\documentclass{article}

\usepackage[final]{neurips_2023}

\usepackage[utf8]{inputenc} 
\usepackage[T1]{fontenc}    
\usepackage{hyperref}       
\usepackage{url}            
\usepackage{booktabs}       
\usepackage{amsfonts}       
\usepackage{nicefrac}       
\usepackage{microtype}      
\usepackage{xcolor}         
\usepackage{graphicx}
\usepackage{caption}
\usepackage{subcaption}

\title{Domain Adaptive Graph Neural Networks for Constraining Cosmological Parameters Across Multiple Data Sets}

\author{%
  Andrea Roncoli\\
  Department of Computer Science\\ 
  University of Pisa\\ 
  Pisa, IT 56127 \\
  \texttt{a.roncoli@studenti.unipi.it} \\
  \And
  Aleksandra \'Ciprijanovi\'c \\
  Computational Science and AI Directorate\\ 
  Fermi National Accelerator Laboratory\\ 
  Batavia, IL 60510 \\
  Department of Astronomy and Astrophysics\\
  University of Chicago\\ 
  Chicago, IL 60637\\
  \texttt{aleksand@fnal.gov}
  \And
  Maggie Voetberg \\
  Computational Science and AI Directorate\\
  Fermi National Accelerator Laboratory\\
  Batavia, IL 60510 \\
  \texttt{maggiev@fnal.gov}
  \And
  Francisco Villaescusa-Navarro\\
  Center for Computational Astrophysics\\
  Flatiron Institute \\
  New York, NY 10010 \\
  \texttt{fvillaescusa@flatironinstitute.org}
  \And
  Brian Nord \\
  Computational Science and AI Directorate\\
    Fermi National Accelerator Laboratory\\
    Batavia, IL 60510\\
    Department of Astronomy and Astrophysics\\
    University of Chicago\\
    Chicago, IL 60637\\
    Kavli Institute for Cosmological Physics\\ University of Chicago\\
    Chicago, IL 60637 \\
  \texttt{nord@fnal.gov}
}

\begin{document}

\maketitle

\begin{abstract}

Deep learning models have been shown to outperform methods that rely on summary statistics, like the power spectrum, in extracting information from complex cosmological data sets.
However, due to differences in the subgrid physics implementation and numerical approximations across different simulation suites, models trained on data from one cosmological simulation show a drop in performance when tested on another.
Similarly, models trained on any of the simulations would also likely experience a drop in performance when applied to observational data.
Training on data from two different suites of the CAMELS hydrodynamic cosmological simulations, we examine the generalization capabilities of Domain Adaptive Graph Neural Networks (DA-GNNs). 
By utilizing GNNs, we capitalize on their capacity to capture structured scale-free cosmological information from galaxy distributions. 
Moreover, by including unsupervised domain adaptation via Maximum Mean Discrepancy (MMD), we enable our models to extract domain-invariant features.
We demonstrate that DA-GNN achieves higher accuracy and robustness on cross-dataset tasks (up to $28\%$ better relative error and up to almost an order of magnitude better $\chi^2$). Using data visualizations, we show the effects of domain adaptation on proper latent space data alignment. 
This shows that DA-GNNs are a promising method for extracting domain-independent cosmological information, a vital step toward robust deep learning for real cosmic survey data.

\end{abstract}

\section{Introduction}
\label{sec:intro}
Accurate determination of cosmological parameters using big data from astronomical surveys is a task of paramount importance in modern science. Historically, the extraction of valuable cosmological information has relied on computing summary statistics [31, 16, 15]. More recently, deep learning methods, such as 2D and 3D Convolutional Neural Networks (CNNs), showed great promise in extracting rich non-linear information that summary statistics struggle to capture [32, 39, 29].  
However, CNNs lack scale-invariance, as their analysis is firmly anchored to the grid size of the convolutional kernels, while any information on scales below that is lost. Choosing a superfine grid to avoid information loss, though, would simply yield almost entirely zeros in case of sparse and irregular data, such as galaxy clusterings. Thus, CNNs result in an inadequate method for structured sparse data.
In contrast, Graph Neural Networks (GNNs) [23, 4, 49, 46] can handle structured cosmic web data in a scale-free manner [41, 14]. As with any other model, the typical procedure is to train GNNs on labeled data (like simulations) and then infer cosmological parameters from unlabeled data (like observations). 
However, there is a significant risk of these models not generalizing in the presence of the domain shift between simulations and observations. Systematic biases have been demonstrated even in experiments that train and test on simulations with different subgrid physics [41]. 
Domain adaptation (DA) techniques [11, 43, 18, 27] can be used to increase model robustness to this type of domain shift. Here we propose the use of Domain Adaptive Graph Neural Networks (DA-GNNs) and investigate the utility of distance-based DA losses i.e., Maximum Mean Discrepancy (MMD) [6]. MMD is an unsupervised DA technique because it does not require labeled data, which is paramount for future applications on observations.
We show that our domain-adaptive models achieve stronger generalization across datasets than regular GNN models. Our work is a significant step towards building future models trained on simulations, yet robust enough to work on observational data.

\textbf{Related Work } GNNs have shown great potential for extracting information from large sparse datasets, such as the distribution of galaxies, galaxy clusters, and cosmic large-scale structures [25, 28, 41, 33, 42, 14]. Unfortunately, due to the complexity of most deep learning models, they often learn dataset-specific features, which renders them useless when testing on a different dataset (different simulations or astronomical observations). In astronomy, it has been shown that DA techniques applied to different types of CNNs can substantially improve model performance in cross-dataset applications [8, 10, 9, 37, 21, 2]. Recently, it has been shown that DA can be used on other types of deep learning algorithms such as GNNs [12, 24, 45, 47, 7, 44, 17]. However, DA on GNNs has not been used for any astrophysics or cosmology applications.

\section{Data and Methods}
\label{sec:data}
\textbf{Data} We use galaxy catalogs from the CAMELS [38]] magneto-hydrodynamic simulations, which follow the evolution of dark matter particles and fluid elements (baryons) from redshift $z = 127$ to $z = 0$.
We use snapshots at $z = 0$ from two different simulation suites: 1) IllusrisTNG [30] was generated with Arepo2~\footnote{\url{https://arepo-code.org/}} and employs the IllustrisTNG subgrid physics model; 2) SIMBA [13] was generated with Gizmo3~\footnote{\url{http://www.tapir.caltech.edu/~phopkins/Site/GIZMO.html}} and employs the SIMBA subgrid physics model.
Using two independent models and codebases to simulate galaxies, cosmic gas, and large-scale structures is critical to assess the generalization potential of the machine learning models.
In particular, we use the LH set of both suites, which contains 1000 simulations evolved with different random seeds and different values of two cosmological parameters (total matter density $\Omega_{m}$ and the amplitude of density fluctuations $\sigma_{8}$) and four astrophysical parameters ($A_{SN1}, A_{SN2}, A_{AGN1}, A_{AGN2}$ related to supernovae efficiency and active galactic nuclei (AGN) feedback, respectively)\footnote{CAMELS dataset documentation:  \url{https://camels.readthedocs.io/en/latest/index.html}}.
We use the following features from the galaxy catalogs as input to our models: 3D positions, stellar mass, stellar radius, stellar metallicity, and maximum circular velocity. 

\label{sec:methods}
\textbf{Methods} Following [41], we generate graphs from 3D galaxy catalogs; these graphs are rotation and translation invariant with respect to the catalogs themselves. We later feed them as inputs to the DA-GNN, using the same architecture as in [41], to allow for fair comparison of the results, with the addition of DA techniques. The model is composed of two parts. The first part is a graph encoder that transforms the graphs into a vector in the latent space through graph blocks [4].
The second part is a simple feedforward network that performs regression, predicting the posterior mean $\mu$ and standard deviation $\sigma$ of the $\Omega_{m}$ cosmological parameter. This can be achieved by minimizing the following loss [26, 40]:
\begin{equation} \label{eq:1}
    \mathcal{L}_{\mu,\sigma} = \log(\sum_{i \in batch}(\Omega_{m,i} - \mu_i)^2) + \log(\sum_{i \in batch}((\Omega_{m,i} - \mu_i)^2 - \sigma_i^2)^2), 
\end{equation}
where $\Omega_{m,i}$ is the ground-truth value for the $i$-th sample in the training set batch, and $\mu_i$ and $\sigma_i$ are the mean and standard deviation, respectively, predicted for sample $i$.

\subsection{Domain Adaptation}
Our objective is to create models that generalize across domains i.e., cosmology simulations with different subgrid physics implementations. To assess this, we train on IllustrisTNG and test on SIMBA -- and vice versa. We experiment with the use of MMD, a distance-based DA technique. MMD measures the distance of two probability distributions, based on the notion of embedding probabilities in a reproducing kernel Hilbert space. We include an MMD-based component in the network loss function, following [8, 48]. For two distributions $Z^1$ and $Z^2$ (with $N$ samples each), this is calculated as:

\begin{equation} \label{eq:2}
     \mathcal{L}_{MMD} = \log(\frac{1}{N-1} \sum_{i\neq j}^N[k(z^1_i, z^1_j) + k(z^2_i , z^2_j) - k(z^1_i,z^2_j) - k(z^2_i,z^1_j)]),
\end{equation}

where $k$ is the Gaussian Radial Basis Function kernel and $z^p_q$ is the sample $q$ of distribution $p$ ($Z^1$ or $Z^2$) [6, 34, 22, 48, 8]. The loss is calculated on the latent space distributions produced by the graph encoder when processing samples from SIMBA and IllustrisTNG sets. Our final objective function is
   $\mathcal{L} = \mathcal{L}_{\mu,\sigma} + \lambda \mathcal{L}_{MMD}$,
where $\lambda \geq 0$ controls the relative contribution of the MMD loss and is a hyperparameter of the model. We find that $\lambda \approx 0.1$ for the best-performing models in this work. The MMD component of the total loss causes the graph encoder to generate similar latent distributions for both simulations, which will improve the performance of the regressor on cross-dataset tasks.

\textbf{Optimization and Computing Resources} We performed experiments on NVIDIA A100 40GB GPU. For each of the models, implemented using PyTorch Geometric [19], we perform a hyperparameter search using the Optuna library [1], with 50 trials per model. More details on code performance, model implementations, and selected hyperparameters can be found in the publicly available code\footnote{https://github.com/deepskies/GNN\_DomainAdapt}.

\subsection{Evaluation}
We split both IllustrisTNG and SIMBA data into training/validation/testing sets with a proportion of 70\%/15\%/15\%. During training, we save the final models at the epoch with the best validation score. For performance metrics, we use the mean relative error $\epsilon$ (reported in percentages), the coefficient of determination $R^2$, and the $\chi^2$ ($N$ = 150 test points), measured as:

\begin{equation} \label{eq:4}
    \epsilon = \frac{1}{N} \sum_{i=1}^{N} \frac{|\Omega_{m,i} -  \mu_i|}{\Omega_{m,i}}, \ \   R^2 = 1 - \frac{\sum_{i=1}^{N} (\Omega_{m,i} - \mu_i)^2}{\sum_{i=1}^{N} (\Omega_{m,i} - \overline{\Omega}_{m})^2}, \ \  \chi^2 = \frac{1}{N} \sum_{i=1}^{N} \frac{(\Omega_{m,i} - \mu_i)^2}{\sigma_i^2},
\end{equation}

where $\overline{\Omega}_{m}$ is the mean of ${\Omega}_{m}$  value in the test set.
A value of $\chi^2$ close to 1 suggests that the standard deviations are correctly predicted and can be seen as minimizing the second term of Equation 1. A higher (lower) value can be seen as an underestimation (overestimation) of the uncertainties [3].

\section{Results}
\label{sec:results}
DA-GNN achieves significantly better results (up to $28\%$ better relative error $\epsilon$ and up to almost an order of magnitude better $\chi^2$) on cross-domain generalization with respect to CosmoGraphNet, whilst achieving comparable results on the same domain test set~\footnote{In [41], authors get slightly better results for the same domain, and slightly worse for the cross-domain tests. We impute these differences to choices such as batch sizes and optimization techniques we took due to computational and time constraints.}, as shown in Table 1 and Figure 1. In [39], the authors were able to infer the value of $\Omega_{m}$ with higher cross-domain accuracy.
However, that analysis utilizes the full matter surface density maps i.e., 2D images, instead of the full 3D galaxy distributions. 
In [14], the authors propose a GNN-based model that performs well cross-domain when trained on the Astrid simulation [5] alone. However, this apparent robustness is achieved by choosing Astrid as the training set and by using input features that are less subject to simulation code variability -- galaxy positions and 1D velocities. When authors try training on other simulations or using more simulation-dependant parameters (e.g., stellar mass), cross-dataset performance drops significantly. Therefore, domain-shift robustness across different cosmological datasets requires DA.

\begin{table}
  \caption{Comparison of results: No Domain Adaptation (top) and MMD (bottom).}
  \label{tab:results}
  \centering
  \scalebox{0.88}{
  \begin{tabular}{lllllllllllll}
    \toprule
    & \multicolumn{3}{c}{I -> I} & \multicolumn{3}{c}{I -> S} & \multicolumn{3}{c}{S -> S} & \multicolumn{3}{c}{S -> I} \\
    \cmidrule(r){2-4} \cmidrule(r){5-7} \cmidrule(r){8-10} \cmidrule(r){11-13}
     & $R^2$ & $\epsilon$ & $\chi^2$ & $R^2$ & $\epsilon$ & $\chi^2$& $R^2$ & $\epsilon$ & $\chi^2$ & $R^2$ & $\epsilon$ & $\chi^2$\\
    \midrule
    NoDA  & 0.97 & 5.0 & 1.39 & -1.04 & 43.8 & 59.43 & 0.97 & 5.2 & 1.79 & 0.22 & 25.0 & 185.54 \\
    MMD & 0.97 & 4.7 & 1.12 & \textbf{0.69} & \textbf{15.7} & \textbf{17.99} & 0.97 & 5.9 & 1.54 & \textbf{0.68} & \textbf{16.7} & \textbf{19.96} \\
    \bottomrule
  \end{tabular}}
\end{table}

\begin{figure}[h]
\centering
\subfloat{
  \includegraphics[width=43mm]{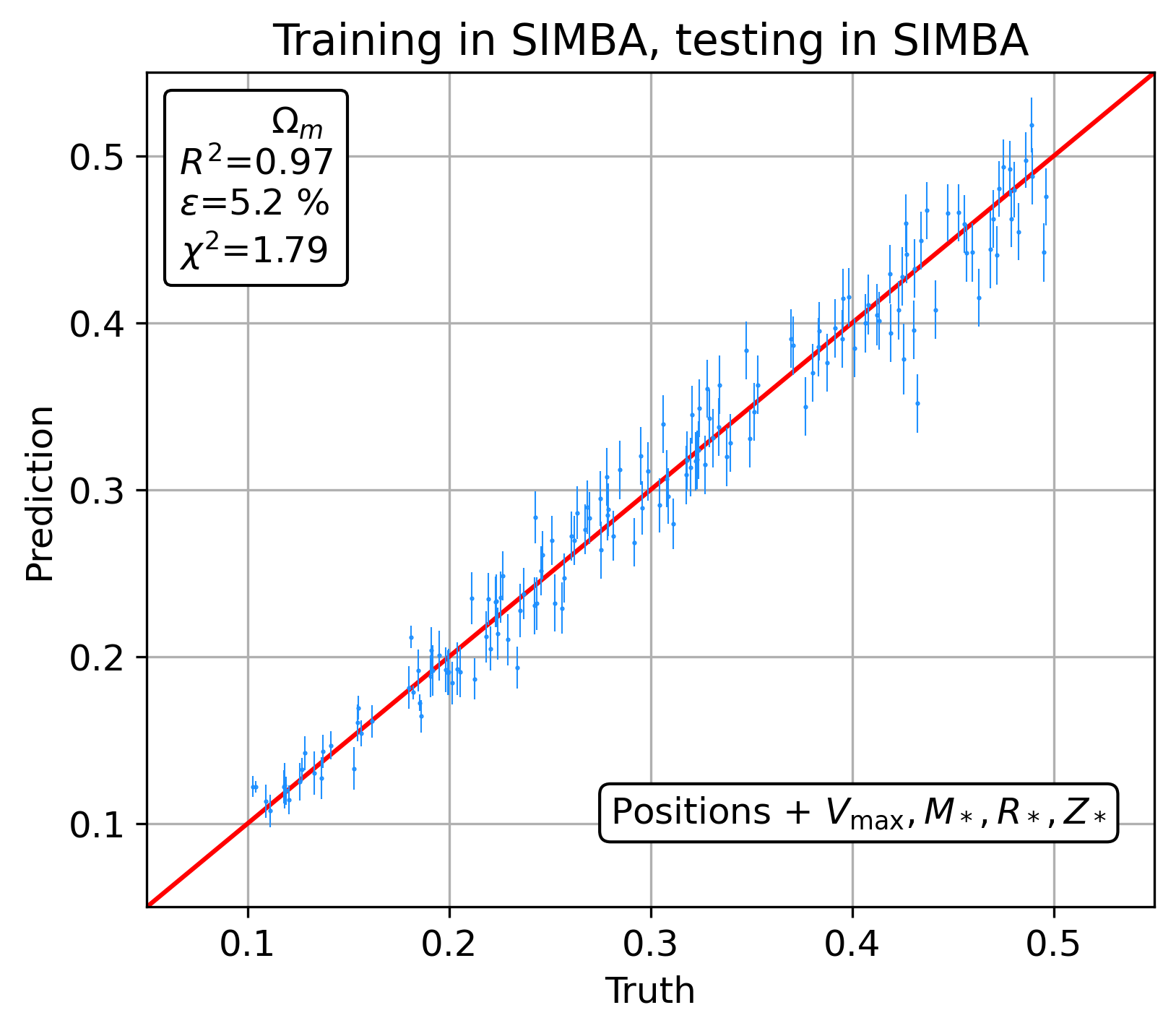}
}
\subfloat{
  \includegraphics[width=43mm]{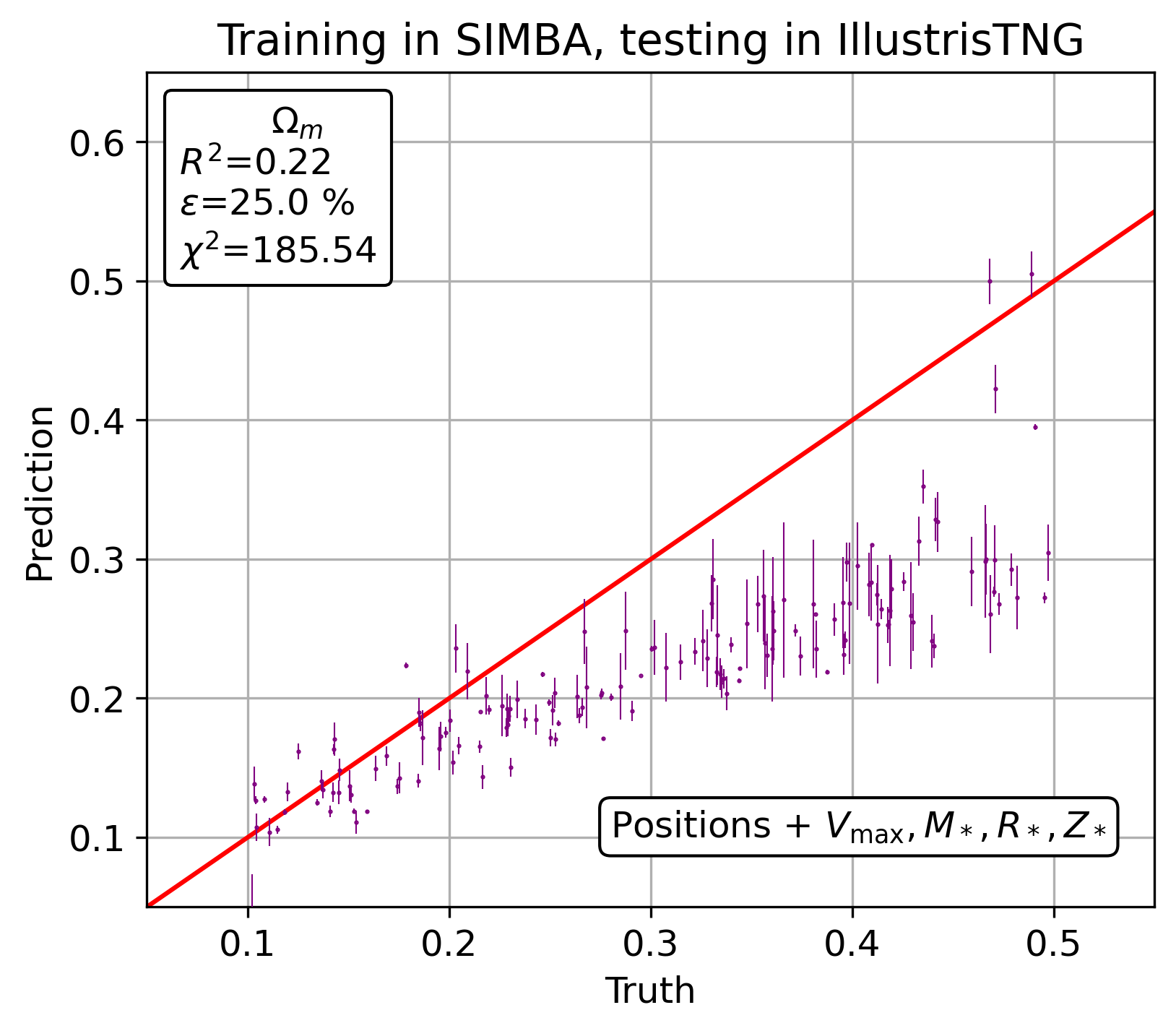}
}
\subfloat{   
  \includegraphics[width=44mm]{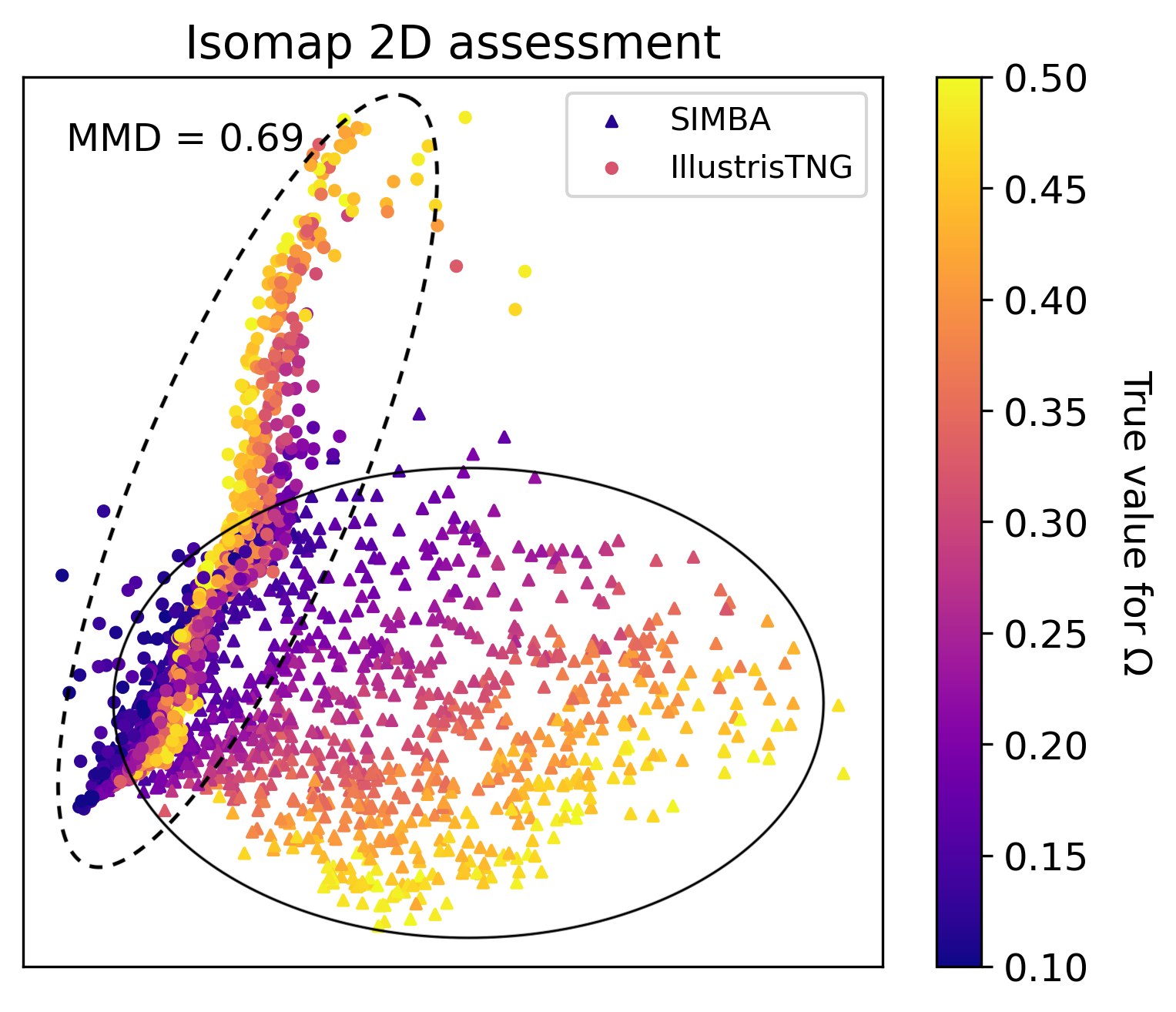}
}

\hspace{0mm}

\subfloat{
  \includegraphics[width=43mm]{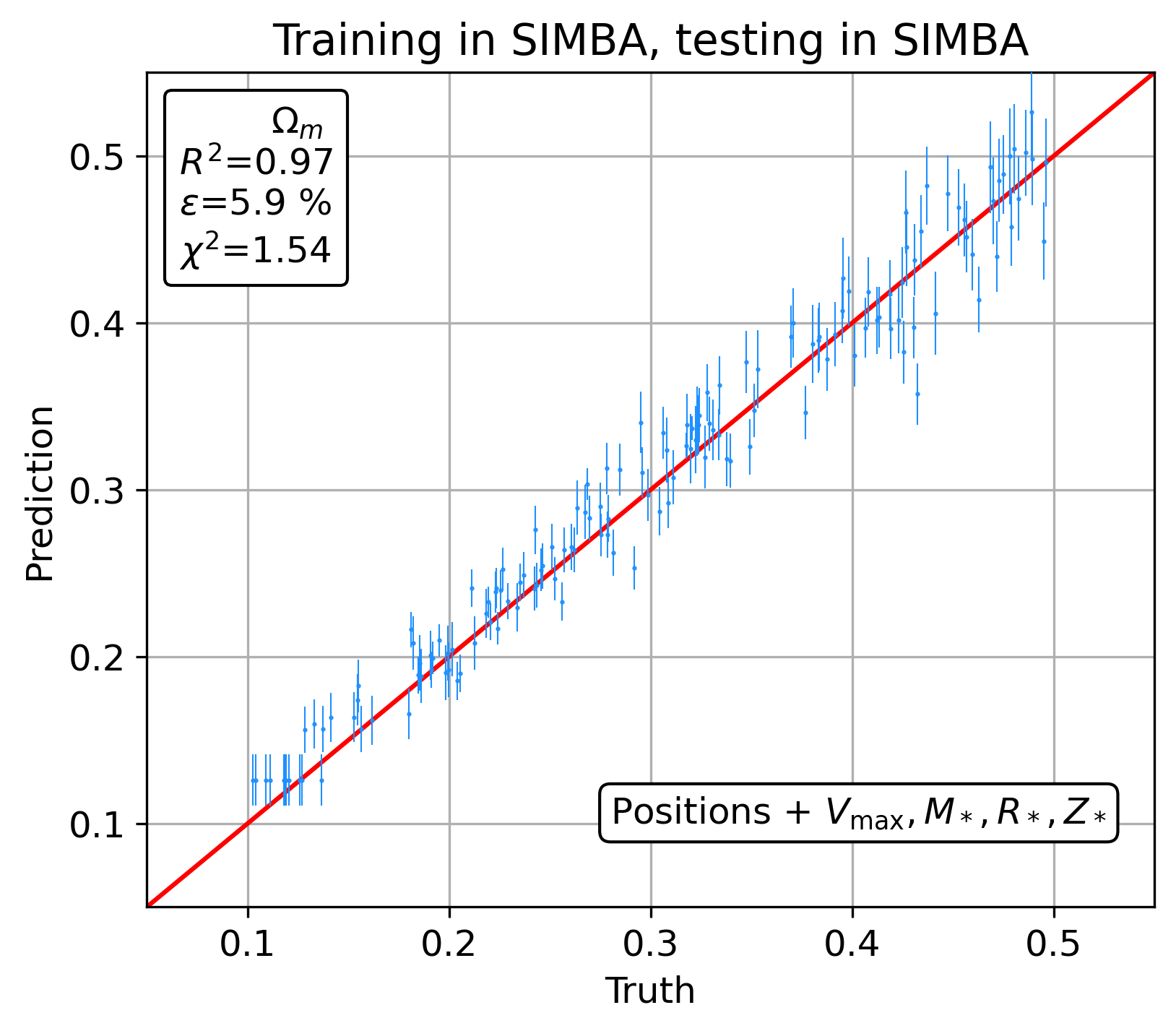}
}
\subfloat{
  \includegraphics[width=43mm]{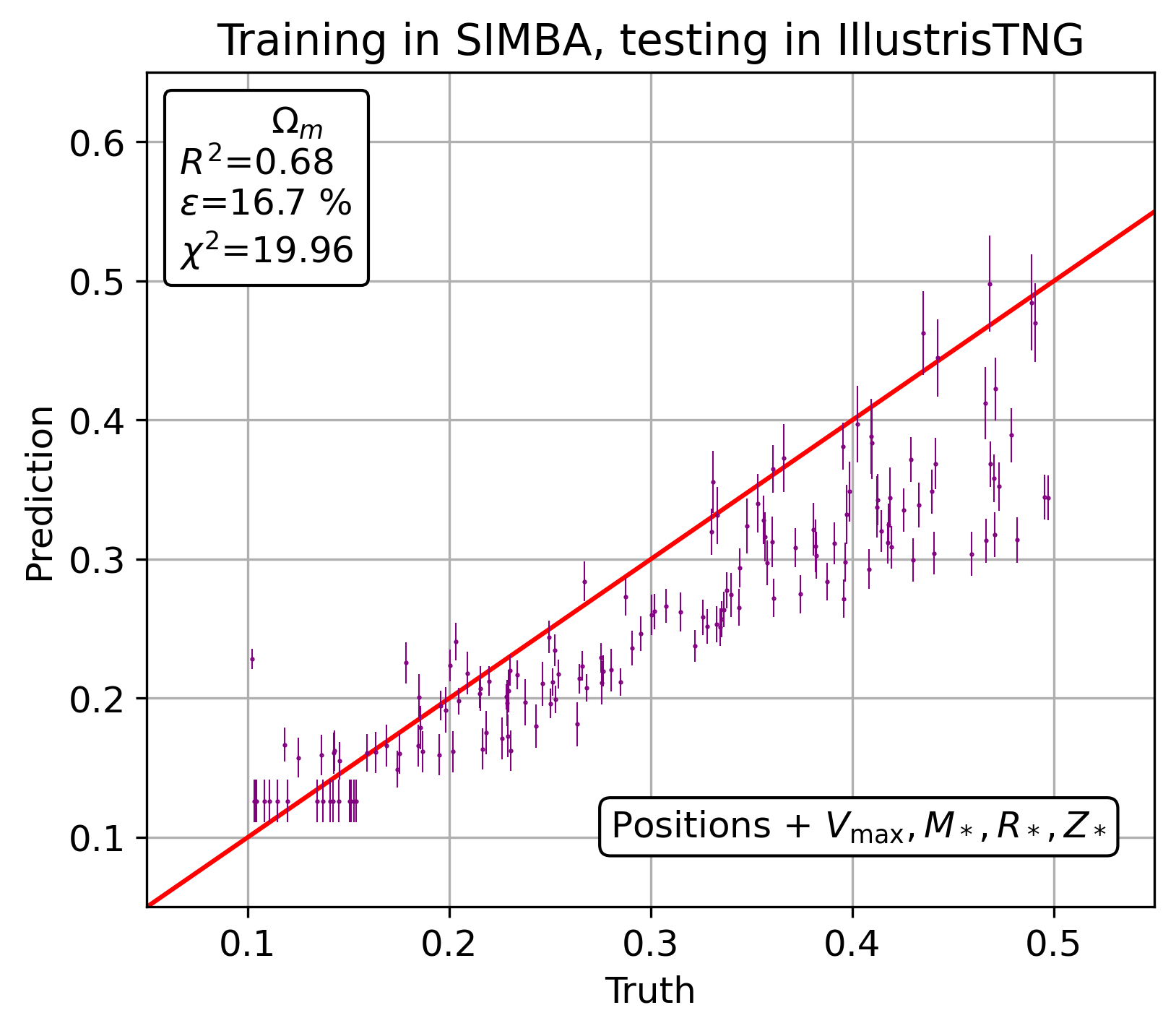}
}
\subfloat{
  \includegraphics[width=44mm]{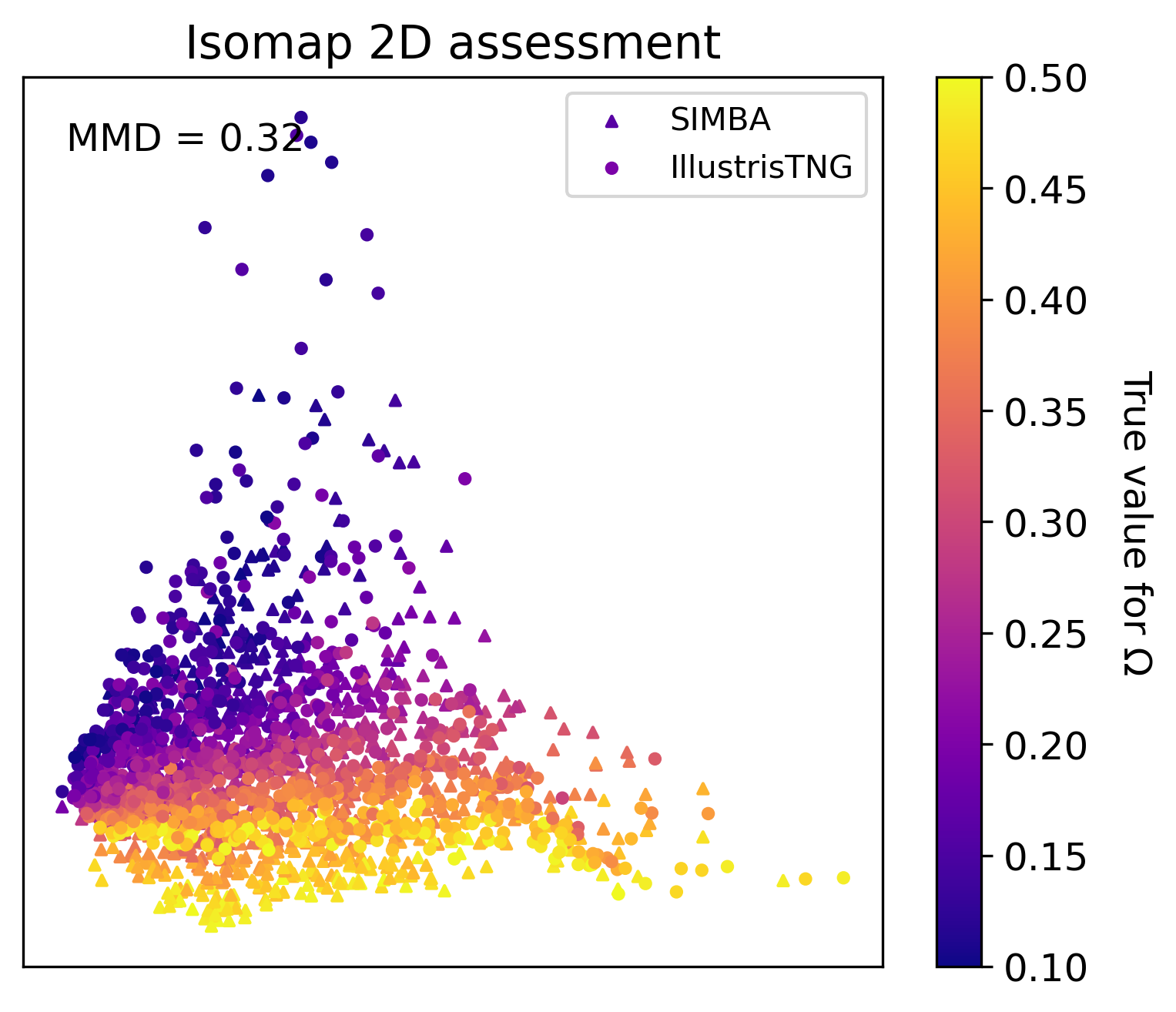}
}

\caption{Comparison of models without (top row) and with DA (bottom row), trained on the SIMBA suite. Training data graphs include 3D positions, maximum circular velocity $V_\mathrm{max}$, stellar mass $M_\star$, stellar radius $R_\star$, and stellar metallicity $Z_\star$. From left to right, we report: a scatter plot for the value of $\Omega_m$ on 1) the same domain, 2) cross-domain and 3) the isomap showing how the GNN is encoding the two datasets in the latent space (SIMBA - triangles, IllustrisTNG - circles)\protect\footnotemark. In the non-domain adapted isomap, ellipses highlight regions where distributions lie, showing the difference between simulation encodings that leads to a substantial drop in performance on the cross-domain task.}
\label{fig:SIMBA}
\end{figure}
\footnotetext{In Appendix \ref{app:plots}, the IllustrisTNG counterpart of this plot is presented.}

\textbf{Latent space organization } Isomaps are two-dimensional projections of the multi-dimensional latent space [35].  Figure 1 shows the difference in the latent space structure without (top row) and with (bottom row) DA. Ellipses in the top right isomap highlight how the two distributions are encoded in different regions of the latent space. Without the MMD loss, the model encodes samples with very different values of $\Omega_m$ close to each other, if they originate from different simulations (circles and triangles of different colors are overlapping). This scenario leads to the fragility of the regressor, which cannot learn to output different values for the same latent space encodings. 
On the contrary, the DA-GNN (bottom right plot) correctly encodes the samples in a domain-invariant way. Visually, circle and triangle distributions are overlapping, which indicates domain mixing. Furthermore, the direction in the color gradient shows that the DA-GNN encodes information such that the regressor can now more correctly predict cosmological parameters based on the encodings of both simulations.

\section{Conclusions}
We propose and demonstrate a method for unsupervised DA for cosmological inference with GNNs. We use an MMD-based loss to enable the domain-invariant encoding of features by the GNN. This approach enhances cross-domain robustness: compared to previous methods, DA-GNNs reduce prediction error and improve uncertainty estimates.

\textbf{Limitations} The cross-domain accuracy remains worse when compared to single-domain performance. Although reaching the same accuracy might not be possible, more flexible approaches such as adversarial-based DA techniques [20, 36], instead of distance-based ones such as MMD, might yield better results. 
Moreover, due to computational and time constraints, our models have been trained and tested only on two of the four available CAMELS simulation suites. Using more suites would yield better cross-domain efficacy and reliability at assessment time. These limitations will be addressed in future work.

\begin{ack}
This manuscript has been supported by Fermi Research Alliance, LLC under Contract No. DE-AC02-07CH11359 with the U.S. Department of Energy (DOE), Office of Science, Office of High Energy Physics. 

This work was supported by the EU Horizon 2020 Research and Innovation Programme under the Marie Sklodowska-Curie Grant Agreement No. 690835, 734303, 822185, 858199, 101003460.
The CAMELS project is supported by the Simons Foundation and the NSF grant AST2108078.

The authors of this paper have committed themselves to performing this work in an equitable, inclusive, and just environment, and we hold ourselves accountable, believing that the best science is contingent on a good research environment.
We acknowledge the Deep Skies Lab as a community of multi-domain experts and collaborators who have facilitated an environment of open discussion, idea generation, and collaboration. This community was important for the development of this project.

Furthermore, we also thank the anonymous referees who helped improve this manuscript.

\textbf{Author Contributions:} 


A.~Roncoli: \textit{Methodology, Software, Validation, Formal analysis, Investigation, Writing - Original Draft, Visualization}; 

A.~\'Ciprijanovi\'c: \textit{Conceptualization, Methodology, Project administration, Resources, Software, Supervision, Writing - Original Draft, Funding Acquisition}; 

M.~Voetberg: \textit{Software, Writing (review and editing)}; 

F.~Villaescusa-Navarro: \textit{Software, Writing (review and editing)}; 

B.~Nord: \textit{Supervision, Resources, Writing (review and editing)}.

\medskip
\small
\end{ack}

\medskip

\section*{References}

[1] Takuya Akiba, Shotaro Sano, Toshihiko Yanase, Takeru Ohta, and Masanori Koyama. Optuna: A next-generation hyperparameter optimization framework. In Proceedings of the 25th ACM SIGKDD International Conference on Knowledge Discovery and Data Mining, 2019.

[2] Stephon Alexander, Sergei Gleyzer, Hanna Parul, Pranath Reddy, Marcos Tidball, and Michael W. Toomey. Domain Adaptation for Simulation-based Dark Matter Searches with Strong Gravitational Lensing. Astrophysical Journal, 954(1):28, September 2023.

[3] Rene Andrae, Tim Schulze-Hartung, and Peter Melchior. Dos and don’ts of reduced chi-squared. arXiv preprint arXiv:1012.3754, 2010.

[4] Peter W Battaglia, Jessica B Hamrick, Victor Bapst, Alvaro Sanchez-Gonzalez, Vinicius Zambaldi, Mateusz Malinowski, Andrea Tacchetti, David Raposo, Adam Santoro, Ryan Faulkner, et al. Relational inductive biases, deep learning, and graph networks. arXiv preprint arXiv:1806.01261, 2018.

[5] Simeon Bird, Yueying Ni, Tiziana Di Matteo, Rupert Croft, Yu Feng, and Nianyi Chen. The ASTRID simulation: galaxy formation and reionization. Monthly Notices of the Royal Astronomical Society, 512(3):3703–3716, May 2022.

[6] Karsten M Borgwardt, Arthur Gretton, Malte J Rasch, Hans-Peter Kriegel, Bernhard Schölkopf, and Alex J Smola. Integrating structured biological data by kernel maximum mean discrepancy. Bioinformatics, 22(14):e49–e57, 2006.

[7] Ruichu Cai, Fengzhu Wu, Zijian Li, Pengfei Wei, Lingling Yi, and Kun Zhang. Graph domain adaptation: A generative view. arXiv preprint arXiv:2106.07482, 2021.

[8] A. \'Ciprijanovi\'c, D. Kafkes, K. Downey, S. Jenkins, G. N. Perdue, S. Madireddy, T. Johnston, G. F. Snyder, and B. Nord. DeepMerge - II. Building robust deep learning algorithms for merging galaxy identification across domains. Monthly Notices of the Royal Astronomical Society, 506(1):677–691, September 2021.

[9] A. \'Ciprijanovi\'c, A. Lewis, K. Pedro, S. Madireddy, B. Nord, G. N. Perdue, and S. M. Wild. DeepAstroUDA: semi-supervised universal domain adaptation for cross-survey galaxy morphology classification and anomaly detection. Machine Learning: Science and Technology, 4(2):025013, June 2023.

[10] Aleksandra \'Ciprijanovi\'c, Diana Kafkes, Gregory Snyder, F. Javier Sánchez, Gabriel Nathan Perdue, Kevin Pedro, Brian Nord, Sandeep Madireddy, and Stefan M. Wild. DeepAdveraries: examining the robustness of deep learning models for galaxy morphology classification. Machine Learning: Science and Technology, 3(3):035007, September 2022.

[11] Gabriela Csurka. Domain Adaptation for Visual Applications: A Comprehensive Survey. arXiv e-prints, page arXiv:1702.05374, February 2017.

[12] Quanyu Dai, Xiao-Ming Wu, Jiaren Xiao, Xiao Shen, and Dan Wang. Graph transfer learning via adversarial domain adaptation with graph convolution. IEEE Transactions on Knowledge and Data Engineering, 35(5):4908–4922, 2022.

[13] Romeel Dav{\'e}, Daniel Angl{\'e}s-Alc{\'a}zar, Desika Narayanan, Qi Li, Mika H. Rafieferantsoa, and Sarah Appleby. SIMBA: Cosmological simulations with black hole growth and feedback. Monthly Notices of the Royal Astronomical Society, 486(2):2827–2849, June 2019.

[14] Natal{\'\i} S. M. de Santi, Helen Shao, Francisco Villaescusa-Navarro, L. Raul Abramo, Romain Teyssier, Pablo Villanueva-Domingo, Yueying Ni, Daniel {Angl{\'e}s-Alc{\'a}zar}, Shy Genel, Elena {Hern{\'a}ndez-Mart{\'\i}nez}, Ulrich P. Steinwandel, Christopher C. Lovell, Klaus Dolag, Tiago Castro, and Mark Vogelsberger. Robust Field-level Likelihood-free Inference with Galaxies. Astrophysical Journal, 952(1):69, July 2023.

[15] DES and SPT Collaborations, T. M. C. Abbott, M. Aguena, A. Alarcon, O. Alves, A. Amon, F. Andrade-Oliveira, J. Annis, B. Ansarinejad, S. Avila, D. Bacon, and et al. Joint analysis of Dark Energy Survey Year 3 data and CMB lensing from SPT and Planck. III. Combined cosmological constraints. Phys. Rev. D, 107(2):023531, January 2023.

[16] DES Collaboration, T. M. C. Abbott, M. Aguena, A. Alarcon, S. Allam, O. Alves, A. Amon, F. Andrade-Oliveira, J. Annis, S. Avila, D. Bacon, and et. al. Dark Energy Survey Year 3 results: Cosmological constraints from galaxy clustering and weak lensing. Phys. Rev. D, 105(2):023520, January 2022.

[17] Zhengming Ding, Sheng Li, Ming Shao, and Yun Fu. Graph adaptive knowledge transfer for unsupervised domain adaptation. In Proceedings of the European Conference on Computer Vision (ECCV), pages 37–52, 2018.

[18] Abolfazl Farahani, Sahar Voghoei, Khaled Rasheed, and Hamid R Arabnia. A brief review of domain adaptation. Advances in data science and information engineering: proceedings from ICDATA 2020 and IKE 2020, pages 877–894, 2021.

[19] Matthias Fey and Jan E. Lenssen. Fast graph representation learning with PyTorch Geometric. In ICLR Workshop on Representation Learning on Graphs and Manifolds, 2019.

[20] Yaroslav Ganin, Evgeniya Ustinova, Hana Ajakan, Pascal Germain, Hugo Larochelle, François Laviolette, Mario Marchand, and Victor Lempitsky. Domain-adversarial training of neural networks. The journal of machine learning research, 17(1):2096–2030, 2016.

[21] Sankalp Gilda, Antoine de Mathelin, Sabine Bellstedt, and Guillaume Richard. Unsupervised Domain Adaptation for Constraining Star Formation Histories. arXiv e-prints, page arXiv:2112.14072, December 2021.

[22] Arthur Gretton, Karsten M Borgwardt, Malte J Rasch, Bernhard Schölkopf, and Alexander Smola. A kernel two-sample test. The Journal of Machine Learning Research, 13(1):723–773, 2012.

[23] Will Hamilton, Zhitao Ying, and Jure Leskovec. Inductive representation learning on large graphs. Advances in neural information processing systems, 30, 2017.

[24] Lukas Hedegaard Morsing, Omar Ali Sheikh-Omar, and Alexandros Iosifidis. Supervised Domain Adaptation using Graph Embedding. arXiv e-prints, page arXiv:2003.04063, March 2020.

[25] Yesukhei Jagvaral, François Lanusse, Sukhdeep Singh, Rachel Mandelbaum, Siamak Ravan- bakhsh, and Duncan Campbell. Galaxies and haloes on graph neural networks: Deep generative modelling scalar and vector quantities for intrinsic alignment. Monthly Notices of the Royal Astronomical Society, 516(2):2406–2419, October 2022.

[26] Niall Jeffrey and Benjamin D Wandelt. Solving high-dimensional parameter inference: marginal posterior densities \& moment networks. arXiv preprint arXiv:2011.05991, 2020.

[27] Suruchi Kumari and Pravendra Singh. Deep learning for unsupervised domain adaptation in medical imaging: Recent advancements and future perspectives. arXiv e-prints, page arXiv:2308.01265, July 2023.

[28] T. Lucas Makinen, Tom Charnock, Pablo Lemos, Natalia Porqueres, Alan F. Heavens, and Benjamin D. Wandelt. The Cosmic Graph: Optimal Information Extraction from Large-Scale Structure using Catalogues. The Open Journal of Astrophysics, 5(1):18, December 2022.

[29] Michelle Ntampaka, Daniel J Eisenstein, Sihan Yuan, and Lehman H Garrison. A hybrid deep learning approach to cosmological constraints from galaxy redshift surveys. The Astrophysical Journal, 889(2):151, 2020.

[30] Annalisa Pillepich, Volker Springel, Dylan Nelson, Shy Genel, Jill Naiman, Rüdiger Pakmor, Lars Hernquist, Paul Torrey, Mark Vogelsberger, Rainer Weinberger, and Federico Marinacci. Simulating galaxy formation with the IllustrisTNG model. Monthly Notices of the Royal Astronomical Society, 473(3):4077–4106, January 2018.

[31] Planck Collaboration, N. Aghanim, Y. Akrami, M. Ashdown, J. Aumont, C. Baccigalupi, M. Ballardini, A. J. Banday, R. B. Barreiro, N. Bartolo, S. Basak, and et al. Planck 2018 results. VI. Cosmological parameters. Astron. Astroph., 641:A6, September 2020.

[32] Dezs{\H{o}} Ribli, B{\'a}lint {\'A}rmin Pataki, Jos{\'e} Manuel Zorrilla Matilla, Daniel Hsu, Zolt{\'a}n Haiman,
and Istv{\'a}n Csabai. Weak lensing cosmology with convolutional neural networks on noisy data. Monthly Notices of the Royal Astronomical Society, 490(2):1843–1860, 2019.

[33] Helen Shao, Francisco Villaescusa-Navarro, Pablo Villanueva-Domingo, Romain Teyssier, Lehman H Garrison, Marco Gatti, Derek Inman, Yueying Ni, Ulrich P Steinwandel, Mihir Kulkarni, et al. Robust field-level inference of cosmological parameters with dark matter halos. The Astrophysical Journal, 944(1):27, 2023.

[34] Alex Smola, Arthur Gretton, Le Song, and Bernhard Schölkopf. A hilbert space embedding for distributions. In International conference on algorithmic learning theory, pages 13–31. Springer, 2007.

[35] Joshua B. Tenenbaum, Vin de Silva, and John C. Langford. A Global Geometric Framework for Nonlinear Dimensionality Reduction. Science, 290(5500):2319–2323, December 2000.

[36] Eric Tzeng, Judy Hoffman, Kate Saenko, and Trevor Darrell. Adversarial discriminative domain adaptation. In Proceedings of the IEEE conference on computer vision and pattern recognition, pages 7167–7176, 2017.

[37] Ricardo Vilalta, Kinjal Dhar Gupta, Dainis Boumber, and Mikhail M. Meskhi. A General Approach to Domain Adaptation with Applications in Astronomy. Publ. Astron. Soc. Pac., 131(1004):108008, October 2019.

[38] Francisco Villaescusa-Navarro, Daniel Anglés-Alcázar, Shy Genel, David N. Spergel, Rachel S. Somerville, Romeel Dave, Annalisa Pillepich, Lars Hernquist, Dylan Nelson, Paul Torrey, Desika Narayanan, Yin Li, Oliver Philcox, Valentina La Torre, Ana Maria Delgado, Shirley Ho, Sultan Hassan, Blakesley Burkhart, Digvijay Wadekar, Nicholas Battaglia, Gabriella Contardo, and Greg L. Bryan. The CAMELS Project: Cosmology and Astrophysics with Machine-learning
Simulations. Astrophysical Journal, 915(1):71, July 2021.

[39] Francisco Villaescusa-Navarro, Shy Genel, Daniel Angles-Alcazar, David N Spergel, Yin Li, Benjamin Wandelt, Leander Thiele, Andrina Nicola, Jose Manuel Zorrilla Matilla, Helen Shao, et al. Robust marginalization of baryonic effects for cosmological inference at the field level. arXiv preprint arXiv:2109.10360, 2021.

[40] Francisco Villaescusa-Navarro, Shy Genel, Daniel Angles-Alcazar, Leander Thiele, Romeel Dave, Desika Narayanan, Andrina Nicola, Yin Li, Pablo Villanueva-Domingo, Benjamin Wandelt, et al. The camels multifield data set: Learning the universe’s fundamental parameters with artificial intelligence. The Astrophysical Journal Supplement Series, 259(2):61, 2022.

[41] Pablo Villanueva-Domingo and Francisco Villaescusa-Navarro. Learning Cosmology and Clustering with Cosmic Graphs. Astrophysical Journal, 937(2):115, October 2022.

[42] Pablo Villanueva-Domingo, Francisco Villaescusa-Navarro, Daniel Anglés-Alcázar, Shy Genel, Federico Marinacci, David N Spergel, Lars Hernquist, Mark Vogelsberger, Romeel Dave, and Desika Narayanan. Inferring halo masses with graph neural networks. The Astrophysical Journal, 935(1):30, 2022.

[43] Mei Wang and Weihong Deng. Deep Visual Domain Adaptation: A Survey. arXiv e-prints, page arXiv:1802.03601, February 2018.

[44] Man Wu, Shirui Pan, Chuan Zhou, Xiaojun Chang, and Xingquan Zhu. Unsupervised domain adaptive graph convolutional networks. In Proceedings of The Web Conference 2020, pages 1457–1467, 2020.

[45] Mengxi Wu and Mohammad Rostami. Unsupervised Domain Adaptation for Graph-Structured Data Using Class-Conditional Distribution Alignment. arXiv e-prints, page arXiv:2301.12361, January 2023.

[46] Zonghan Wu, Shirui Pan, Fengwen Chen, Guodong Long, Chengqi Zhang, and S Yu Philip. A comprehensive survey on graph neural networks. IEEE transactions on neural networks and learning systems, 32(1):4–24, 2020.

[47] Nan Yin, Li Shen, Mengzhu Wang, Long Lan, Zeyu Ma, Chong Chen, Xian-Sheng Hua, and Xiao Luo. Coco: A coupled contrastive framework for unsupervised domain adaptive graph classification. arXiv preprint arXiv:2306.04979, 2023.

[48] Wen Zhang and Dongrui Wu. Discriminative joint probability maximum mean discrepancy (djp-mmd) for domain adaptation. In 2020 international joint conference on neural networks (IJCNN), pages 1–8. IEEE, 2020.

[49] Jie Zhou, Ganqu Cui, Shengding Hu, Zhengyan Zhang, Cheng Yang, Zhiyuan Liu, Lifeng Wang, Changcheng Li, and Maosong Sun. Graph neural networks: A review of methods and applications. AI open, 1:57–81, 2020.


\appendix
\newpage
\section{Additional Plots}
\label{app:plots}

\begin{figure}[h!]
\centering
\subfloat{
  \includegraphics[width=43mm]{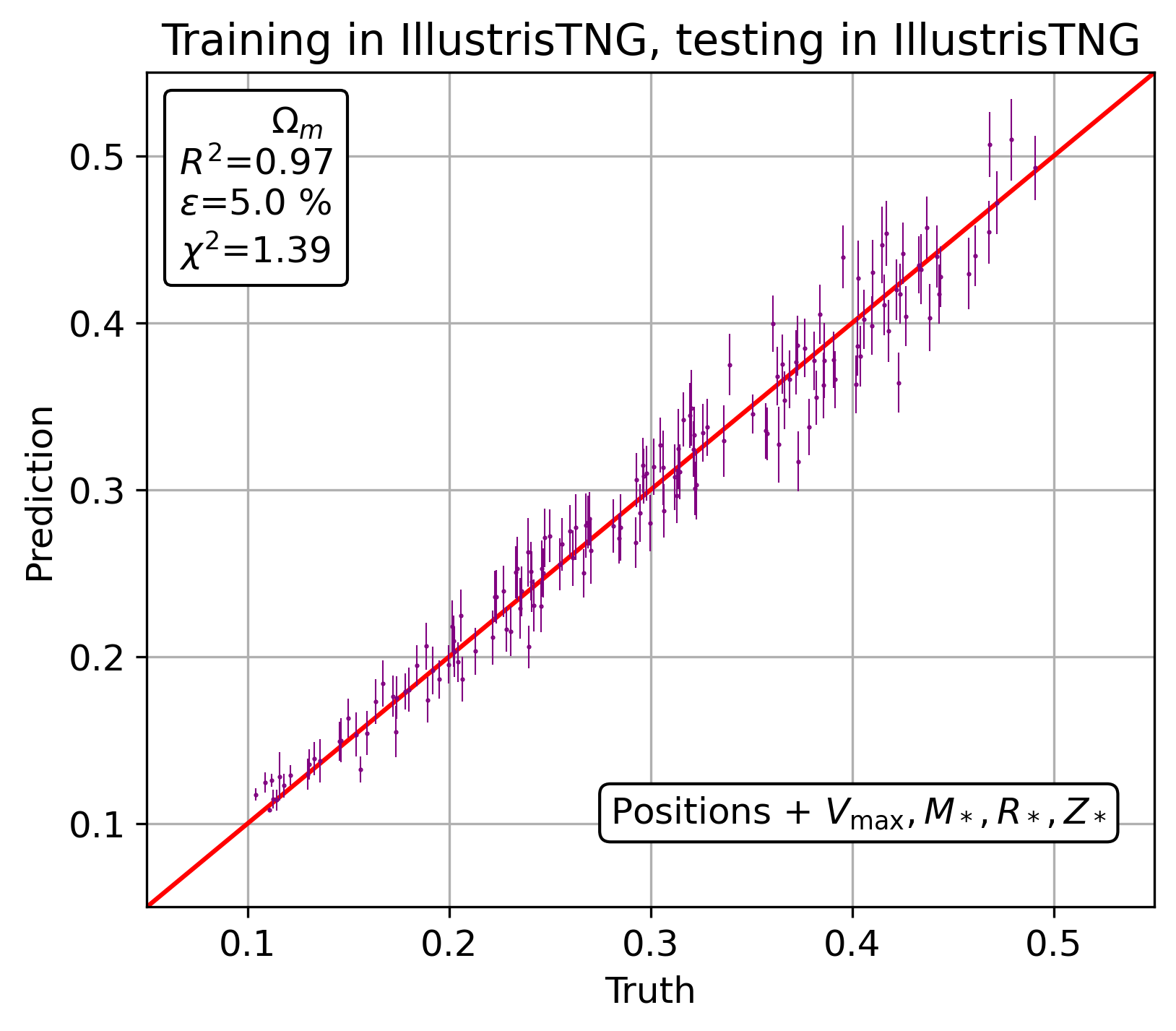}
}
\subfloat{
  \includegraphics[width=43mm]{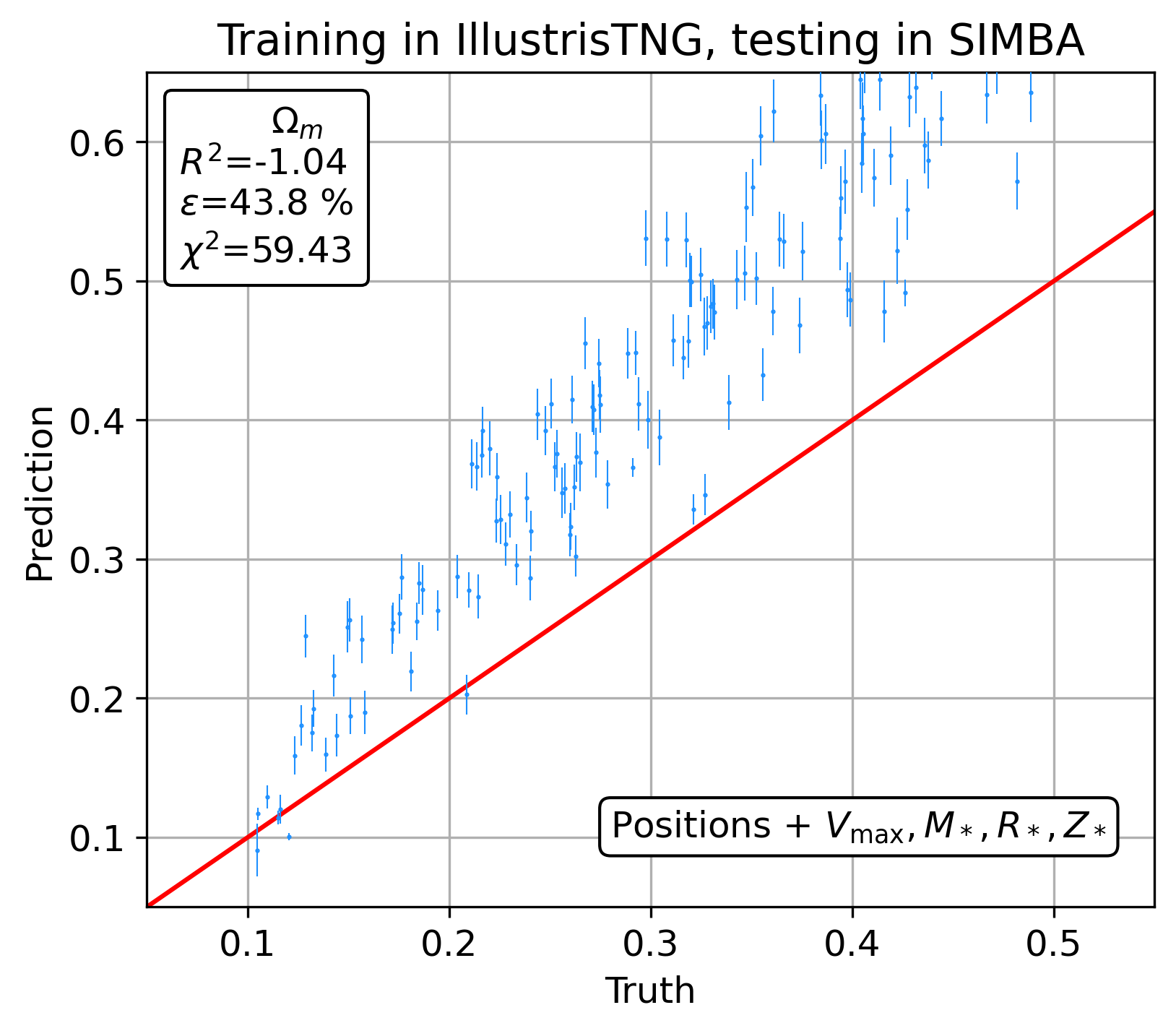}
}
\subfloat{   
  \includegraphics[width=44mm]{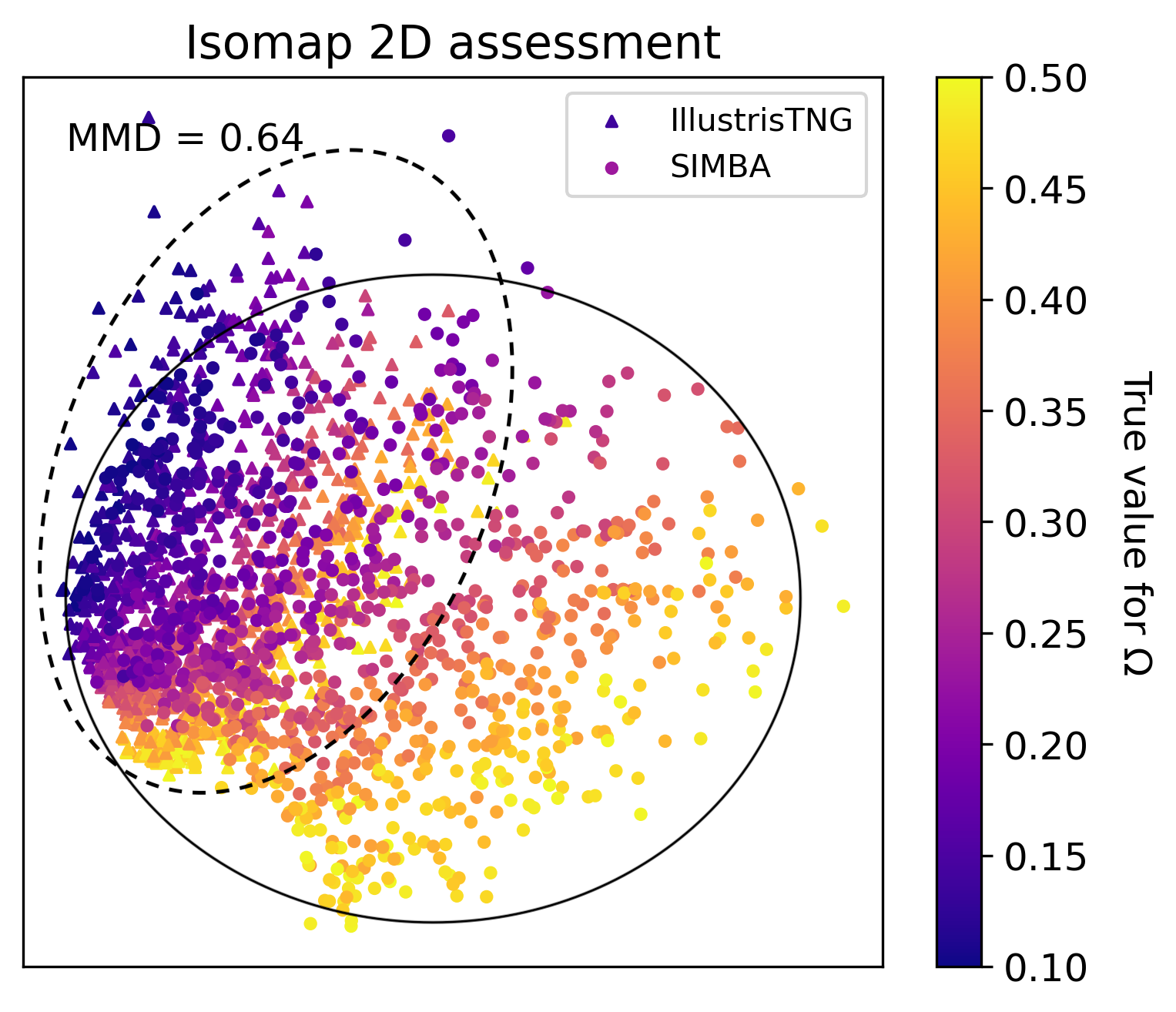}
}

\hspace{0mm}

\subfloat{
  \includegraphics[width=43mm]{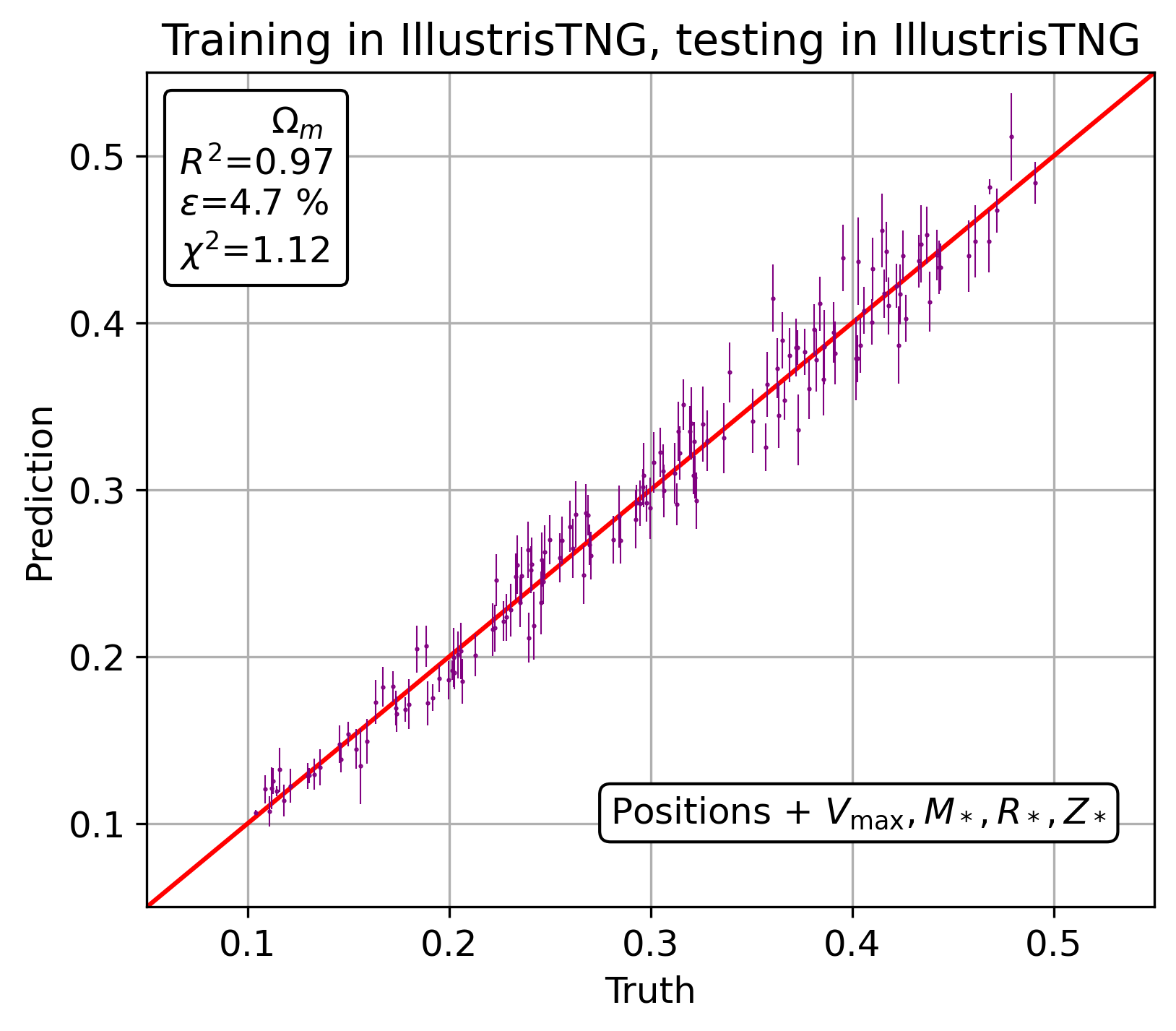}
}
\subfloat{
  \includegraphics[width=43mm]{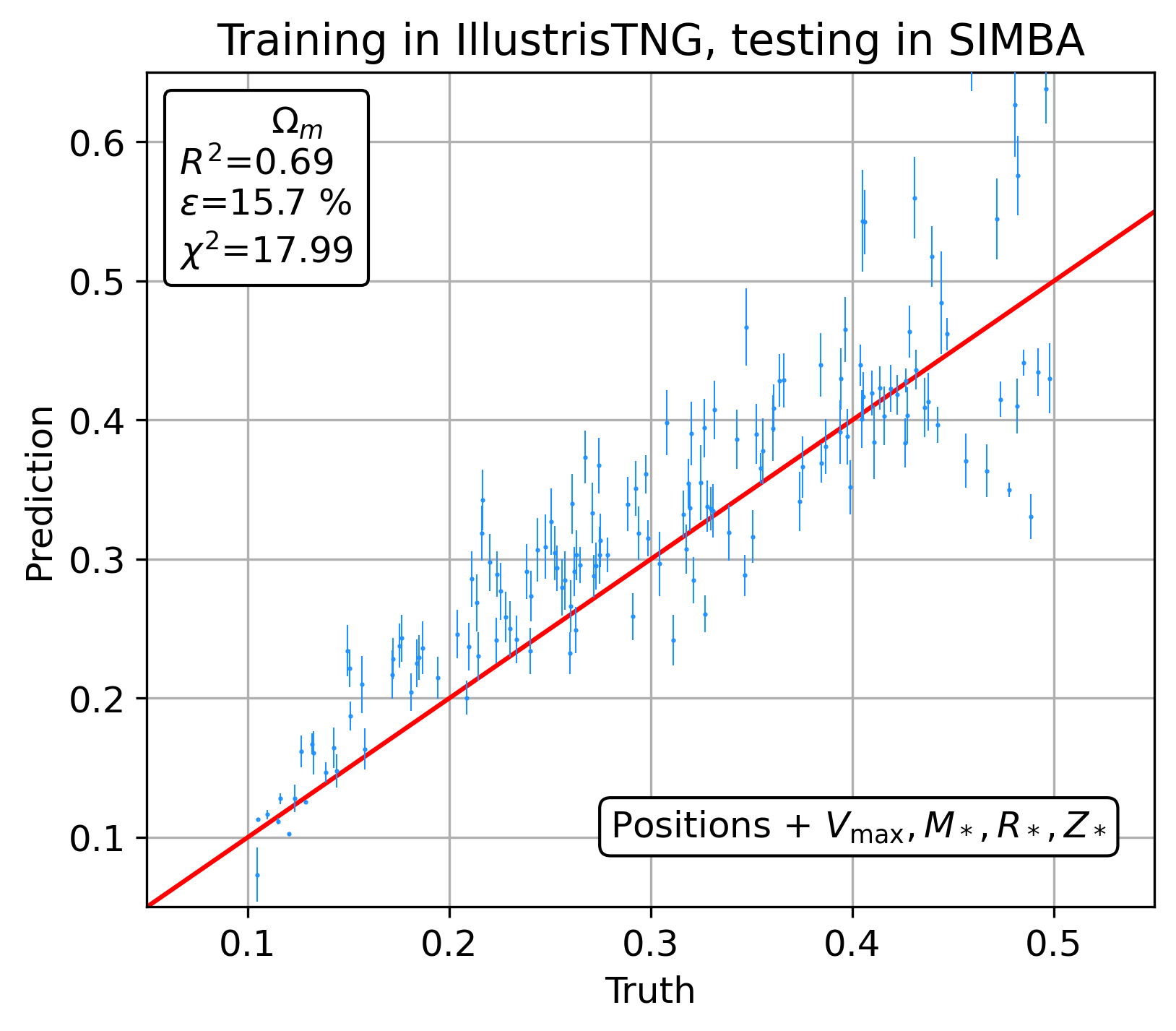}
}
\subfloat{
  \includegraphics[width=44mm]{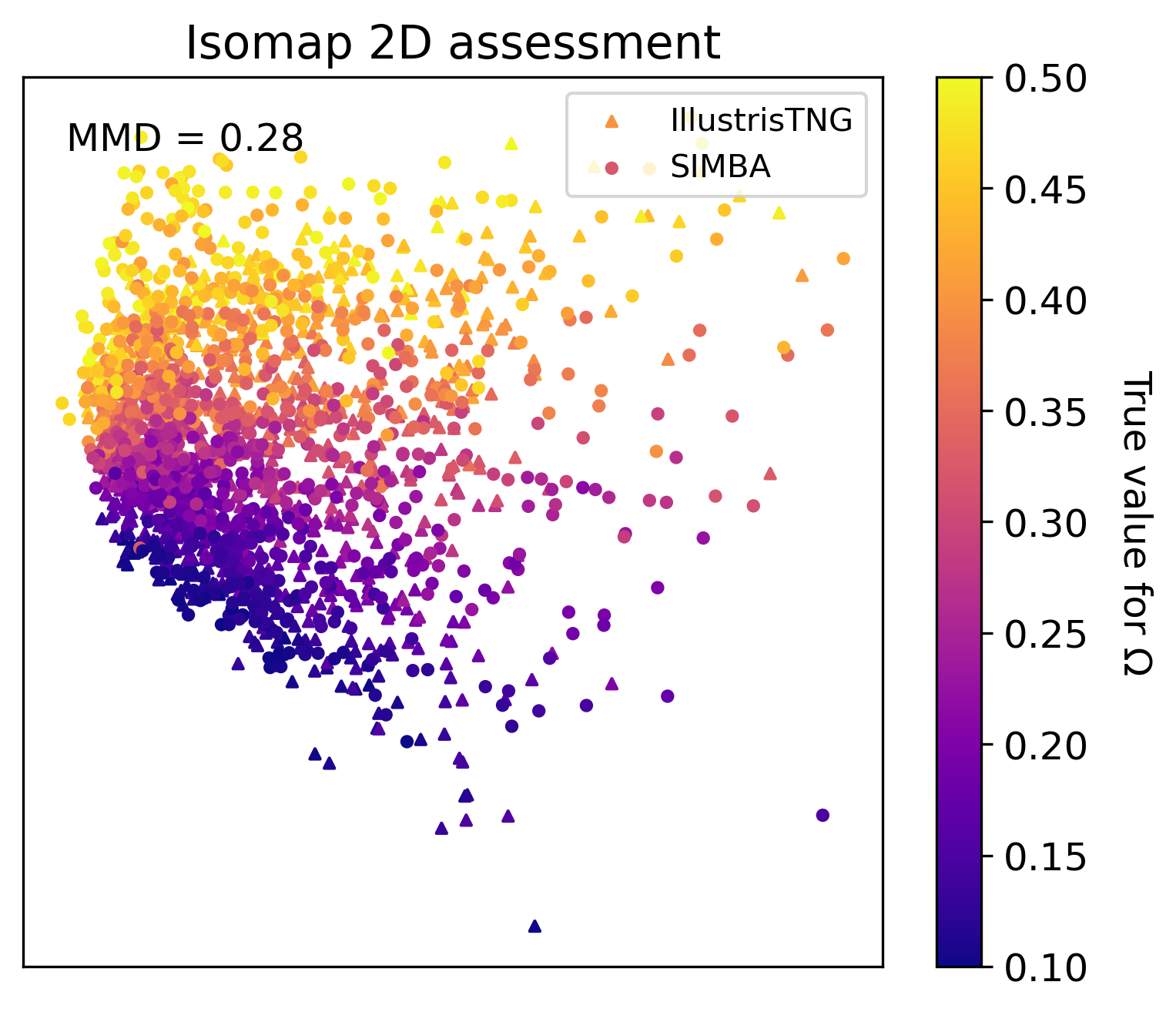}
}

\caption{Comparison of models without (top row) and with DA (bottom row), trained on the IllustrisTNG suite. Training data graphs include 3D positions, maximum circular velocity $V_\mathrm{max}$, stellar mass $M_\star$, stellar radius $R_\star$, and stellar metallicity $Z_\star$. From left to right, we report: a scatter plot for the value of $\Omega_m$ on 1) the same domain, 2) cross-domain and 3) the isomap showing how the GNN is encoding the two datasets in the latent space (IllustrisTNG - triangles, SIMBA - circles). In the non-domain adapted isomap, ellipses highlight regions where distributions lie, showing the difference between simulation encodings that leads to a substantial drop in performance on the cross-domain task.}
\label{fig:Illustris}
\end{figure}
\end{document}